\documentclass{RNTI}
\usepackage[french]{minitoc}
\newtheorem{definition}{Définition}
\usepackage{pifont}
\begin{document}

\thispagestyle{empty}

\fancyhead{} \fancyfoot{} \fancyhead[LE]{
%
%
Hiérarchisation des règles d'association en fouille de textes
%
} \fancyhead[RO]{
%
Bendaoud et al.
%
} \fancyfoot[LE,RO]{RNTI - 1}

\begin{center}
{\Large\bf

Hiérarchisation des règles d'association en fouille de textes

}
~\\~\\

Rokia BENDAOUD$^{*}$, Yannick TOUSSAINT$^{*}$\\
Amedeo NAPOLI$^{*}$\\

~\\

$^{*}$LORIA Campus Scientifique - BP 239\\
54506 VANDOEUVRE-lès-NANCY CEDEX\\
\{bendaoud,toussaint,napoli\}@loria.fr,\\

\end{center}

\begin{abstract}

L'extraction de règles d'association est souvent exploitée comme
méthode de fouille de données. Cependant, une des limites de cette
approche vient du très grand nombre de règles extraites et de la
difficulté pour l'analyste à appréhender la totalité de ces
règles. Nous proposons donc de pallier ce problème en structurant
l'ensemble des règles d'association en hiérarchies. La
structuration des règles se fait à deux niveaux. Un niveau global
qui a pour objectif de construire une hiérarchie structurant  les
règles extraites des données. Nous définissons donc un premier
type de subsomption entre règles issue de la subsomption dans les
treillis de Galois. Le second niveau correspond à une analyse
locale des règles et génère pour une règle donnée une hiérarchie
de généralisation de cette règle qui repose sur des connaissances
complémentaires exprimées dans un modèle terminologique. Ce niveau
fait appel à un second type de subsomption inspiré de la
subsomption en programmation logique inductive.

Nous définissons ces deux types de subsomptions, développons un
exemple montrant l'intérêt de l'approche pour l'analyste et
étudions les propriétés formelles des hiérarchies ainsi proposées.

\end{abstract}

\section{Introduction}

L'extraction des règles d'association appliquée à des textes est
une méthode de fouille de données qui permet de mettre en valeur
des liens entre les termes des textes. Ces liens peuvent alors
être interprétés par des experts en vue, par exemple, de la
construction d'une ontologie.

Que ce soit à partir de textes où à partir de base de données, le
nombre de règles extraites est souvent très grand et difficile à
appréhender par un expert humain. De nombreux travaux se sont
intéressés à élaguer l'ensemble des règles et à les classer soit
par rapport à des critères statistiques, soit par rapport à une
base de connaissances \cite{Janetzko04}. Nous proposons dans cet
article une approche visant à structurer les règles sous forme
hiérarchique afin de permettre à l'expert une approche descendante
de la lecture de l'ensemble des règles. En réalité, nous proposons
à l'expert deux approches d'analyse, un niveau global et un niveau
local, tous deux reposant sur une structuration hiérarchique des
règles. Ces deux types de structuration hiérarchique nous ont
conduit à définir deux types de subsomption qui, au final, peuvent
être combinés.

L'approche globale a pour objectif de permettre à l'expert
d'appréhender l'ensemble des règles extraites à partir d'un
ensemble de textes. L'enjeu est donc de lui proposer une vision
structurée et synthétique de cet ensemble de règles. Nous
considérons alors les termes comme étant non hiérarchisés. La
hiérarchie des règles ainsi construite repose sur un premier type
de subsomption pour laquelle aucune propriété n'est plus générale
(ni plus spécifique) qu'une autre. Par exemple des propriétés
comme $\mathtt{vole, respire, pond}$ peuvent être considérées
comme indépendantes les unes des autres.

Au niveau local, l'expert peut disposé de propriétés structurées.
Nous construisons une hiérarchie locale autour d'une règle ou d'un
sous-ensemble de règles en prenant en compte des propriétés
hiérarchisées au sein d'un modèle terminologique, cette hiérarchie
repose sur un second type de subsomption des règles.

Cet article est divisé en six parties. Dans la section 2, nous
définissons les règles d'association, puis les treillis de Galois
que nous utilisons comme support à l'extraction des règles. La
section 3 précise notre méthodologie d'extraction de règles. La
section 4 présente la structuration globale des règles et définit
le premier type de subsomption qui repose sur la subsomption dans
les treillis. La section suivante décrit la structuration locales
des règles et définit le second type de subsomption permettant de
généraliser les règles, en se reposant sur la subsomption en
programmation logique inductive. Bien que ces travaux soient
inspirés d'une problématique de fouille de texte, les deux types
de subsomption ont été testées sur une base de données réduite
pour faciliter l'interprétation des résultats.
\section{Le contexte mathématique}
\subsection{Les règles d'association}

Pour faire l'analogie avec le vocabulaire généralement utilisé en
fouille de données, nous allons considérer un texte comme un
individu et désignerons par I l'ensemble des individus. Les termes
 sont considérés comme des propriétés et
l'ensemble des propriétés est noté P. Nous considérons la relation
binaire $\mathcal{R}$ tel que $\mathcal{R}$ $\subset$ I $\times$ P
et $\mathcal{R}$ ($\texttt{i,p}$) si l'individu $\texttt{i}$
contient la propriété $\texttt{p}$.

\begin{definition}[Les règles d'association] \mbox{}

Une règle d'association est une implication pondérée de la forme A
$\Rightarrow$ B, où A est la prémisse, et B est la conclusion,
avec A $\subseteq$ P, B $\subseteq$ P et A $\cap$ B = $\emptyset$.
\end{definition}

Les règles d'association \cite{Agrawal93} permettent de mettre en
évidence les dépendances entre les propriétés. Par exemple, la
règle : $\mathtt{vole}$ $\Rightarrow$ $\mathtt{pond}$ $\sqcap$
$\mathtt{respire}$ ($\sqcap$ désigne la conjonction des
propriétés) peut s'interpréter comme le fait que si un individu
$\mathtt{vole}$, il est probable qu'il ait les propriétés
$\mathtt{pond}$ et $\mathtt{respire}$.

\begin{definition}[Motif et image d'un motif] \mbox{}

Un $\textbf{motif}$ est un sous-ensemble de P. On dit qu'un
 individu i contient le motif M, si M et i sont en relation:
$\forall$ p $\in$ M: $\mathcal{R}$(i,p).

L'$\textbf{image d'un motif}$ M est l'ensemble des individus qui
contiennent le motif M.

\end{definition}

Le processus d'extraction des règles d'association est un
processus exponentiel, en fonction du nombre d'individus et du
nombre de propriétés. Il existe plusieurs méthodes pour réduire la
complexité de ce processus, l'une d'elle est l'utilisation des
indices statistique \cite{Cherfi02}. Les deux indices statistiques
les plus couramment utilisés sont le support et la confiance, qui
servent à réduire le nombre de règles extraites.
\paragraph{Rappels : support d'un motif, d'une règle, confiance et motif fréquent}\mbox{}

\textbf{Le support d'un motif} représente le nombre d'individus
qui possèdent le motif sur la cardinalité de l'ensemble des
individus. \vspace{-0.2cm}
\[\vspace{-0.2cm}
support (M_{1})=\frac{Image(M_{1})}{card(I)}\vspace{-0.08cm}
\]

\textbf{Le support d'une règle} représente le nombre d'individus
qui vérifient la règle, c'est-à-dire, qui possèdent le motif
$\mathtt{A \sqcap B}$\vspace{-0.2cm}
\[
support (A \Rightarrow B)=support(A \sqcap B) \vspace{-0.08cm}
\]

\textbf{La confiance}  d'une règle $\mathtt{A}$
$\Rightarrow$ $\mathtt{B}$ est définie par le fait qu'un individu
possède les propriétés B
 sachant qu'il possède celles de A :
\vspace{-0.2cm}
\[\vspace{-0.2cm}
confiance(A \Rightarrow B)= \frac{ Support (A \Rightarrow
B)}{Support(A)}\vspace{-0.08cm}
\]

\textbf{Motif fréquent} Un motif est dit fréquent si et seulement
si son support est supérieur à un seuil $\emph{minsupp}$.

\begin{definition}[Règle valide, règle totale, règle partielle et règle informative] \mbox{}
Soit $\texttt{R : A $\Rightarrow$ B}$ une règle :

La règle R est \textbf{valide} ssi support($\texttt{R}$)$\geq$
minsupp et si sa confiance est supérieure à un seuil
$\texttt{minconf}$.

La règle R est \textbf{totale} ssi confiance($\texttt{R}$) = 1, ce
qui signifie qu'à chaque fois qu'un individu i possède A, i
possède également B. Les règles totales ne possèdent donc pas de
contre-exemple.

La règle R est \textbf{partielle} ssi confiance($\texttt{R}$)$<$1.
Ce sont des règles qui possèdent des contre-exemples, c'est-à-dire
des individus qui possèdent la partie gauche de la règle mais pas
la partie droite.

La règle $\texttt{R}$ est dite \textbf{informative} ssi elle est
valide et A $\cap$ B = $\emptyset$.

\end{definition}

\paragraph{Propriétés des règles}
Ces deux propriétés sont utilisées dans la section \ref{SPLI}.
\begin{description}
    \item[Prop1.] transitivité : si $\mathtt{A}$ $\Rightarrow$ $\mathtt{B}$ et $\mathtt{B}$ $\Rightarrow$ $\mathtt{C}$
    et que l'une des règles est valide et l'autre totales alors $\mathtt{A}$ $\Rightarrow$ $\mathtt{C}$ est valide.\label{P1}
    \item[Prop2.] si $\mathtt{A}$ $\Rightarrow$ $\mathtt{B}$ et que Image($\mathtt{B}$) $\subseteq$ Image($\mathtt{B}$') alors $\mathtt{A}$ $\Rightarrow$
    $\mathtt{B}$'.\label{P2}
\end{description}

Il existe différentes approches pour l'extraction des règles
d'association. La première issue des travaux en bases de données,
est l'extraction de règles à partir des algorithmes de motifs
fréquents. La seconde est l'extraction des règles à partir d'un
treillis de Galois. C'est ce deuxième type d'extraction de règles
que nous allons utiliser.

\subsection{Les treillis de Galois}

Rappelons qu'un treillis de Galois ou treillis de concepts
s'appuie sur une connexion de Galois et organise un ensemble de
concepts formels --les fermés de la connexion-- en un treillis
(\cite{Barbut70}, \cite{Guenoche90}, \cite{Ganter99}). Les
concepts se notent ci-dessous $C_{k}$ = ($P_{k}$,$I_{k}$) où
$P_{k}$ désigne les propriétés du concept $C_{k}$ (l'intension du
concept) et $I_{k}$ les individus recouverts par le concept
(l'extension du concept). La relation d'ordre partiel dans un
treillis vérifie : $C_{k}$$\sqsubseteq$$C_{k'}$ ssi
$I_{k}$$\subseteq$$I_{k'}$ (et de façon duale
$P_{k'}$$\subseteq$$P_{k}$).

\section{Extraction des règles d'association à partir d'un
treillis de Galois}

La formalisation mathématique de l'extraction de règles
d'association à partir d'un treillis de Galois est présentée dans
(\cite{Guigues86}, \cite{Godin95}) et fait appel à la notion de
propriétés propres et de propriétés héritées pour un concept. De
façon analogue, l'extraction de règles que nous proposons se fait
en parcourant les concepts du treillis et en considérant
l'intension du concept comme le motif commun à toutes les règles
extraites à partir de ce concept. Le processus se déroule de la
façon suivante :
\begin{itemize}
    \item Soit $C_{s}$ = ($P_{s}$,$I_{s}$) sommet du treillis.
    \item si le support du motif $P_{s}$ $\geq$ \emph{minsupp}
        \begin{itemize}
        \item alors extraire l'ensemble $R_{s}$ des règles associées au
        motif $P_{s}$ de la forme \\$P_{i}$ $\Rightarrow$
        $P_{s}$$\setminus$$P_{i}$, tel que $P_{i}$ $\subset$ $P_{s}$.
        \item calculer la confiance, supprimer les
        règles donc la confiance $<$ \emph{minconf}.
        \item appeler récursivement l'algorithme pour tous les concepts subsumés par
        $C_{s}$ dans le treillis.
        \end{itemize}
    \item sinon passer à une autre branche du treillis.
     \end{itemize}

Soit la règle $R_{1}$ : \texttt{A $\Rightarrow$ B} extraite du
concept $C_{1}$ = ($P_{1}$,$I_{1}$), nous pouvons calculer
support($R_{1}$) et confiance($R_{1}$) directement du treillis de
Galois

\[\vspace{-0.1cm}
support (R_{1}) = support (P_{1}) = \frac{card(I_{1})}{card(I)}
\hspace{0.2cm} et \hspace{0.2cm} confiance (R_{1}) =
\frac{support(R_{1})}{support(A)}
\]

Pour trouver le support de \texttt{A} qui n'est peut-être pas un
fermé, nous devons chercher le concept dont l'intension est le
fermé minimal contenant le motif \texttt{A}. Pour cela, on part du
sommet du treillis, cherchant le premier concept qui possède dans
son intension le motif \texttt{A}, soit $C_{j}$ = ($P_{j}$
,$I_{j}$) ce concept, alors support(\texttt{A}) =
support($I_{j}$).

Seules les règles issues d'un motif fermé par rapport à la
connexion de Galois sont extraites. L'algorithme est donc plus
restrictif que Apriori (Agrawal et al. 1994) où la notion de fermé
n'est pas utilisée. En revanche, cette méthode ne se limite pas à
l'extraction de règles de type $\texttt{clé $\Rightarrow$
fermé$\setminus$clé}$ comme c'est le cas pour \emph{Close}
\cite{Bastide02}. On obtient donc un sous-ensemble de règles par
rapport à \emph{Close}. \cite{Bastide02} montre que cet ensemble
de règles constitue une base (non minimale). De même, notre
ensemble de règles extraites constitue une base non minimale.
\section{Classification de règles pour des propriétés non hiérarchisées}

La subsomption de règles lorsque les propriétés sont non
hiérarchisées repose directement sur la structure du treillis de
Galois du contexte (I,P,$\mathcal{R}$) introduit en section~2.
Elle est définie à partir de la subsomption sur l'intension des
concepts que nous considérons comme des motifs. Nous appelons
cette subsomption basée sur les motifs M-subsomption et nous la
notons $\sqsubseteq$$_{M}$.

\subsection{Subsomption des règles non hiérarchisées}

\begin{definition}[M-Subsomption des règles]\mbox{}

Soient $C_{1}$ = ($P_{1}$,$I_{1}$) et $C_{2}$ = ($P_{2}$,$I_{2}$)
deux concepts du treillis de Galois. Soient $R_{1}$ une règle
issue du motif $P_{1}$ et $R_{2}$ une règle issue du motif
$P_{2}$.

$R_{2}$ M-subsome $R_{1}$ noté $R_{2}$ $\sqsubseteq$$_{M}$ $R_{1}$
ssi $C_{2}$ $\sqsubseteq$ $C_{1}$ dans le treillis du contexte
(I,P,$\mathcal{R}$).
\end{definition}

\subsection{Les R-ensembles}

\begin{definition}[R-ensemble]\mbox{}

Soit M un motif de longueur $\geq$ 1. Un $\textbf{R-ensemble}$
engendré pour M, noté\linebreak \textbf{R}(M), est défini comme
l'ensemble des règles valides qu'il est possible d'extraire de M.
\end{definition}

Le fait que deux règles soient du même R-ensemble, signifie
qu'elles ont été extraites du même concept dans le treillis. De ce
fait nous allons les placer dans le même noeud de la hiérarchie
des règles. Les règles d'un même R-ensemble ont le même support
(en extension) mais pas forcément la même confiance.

Exemple: Soient le motif $\mathtt{P_1}$=

\{$\mathtt{respire,pond,vole}$\}. Nous pouvons extraire 7 règles, supposons que 3 seulement sont valides:\\
$R_{1}$ : $\mathtt{respire}$$\Rightarrow$$\mathtt{pond,vole}$,
$R_{2}$ : $\mathtt{pond}$$\Rightarrow$$\mathtt{respire,vole}$, $R_{3}$ : $\mathtt{vole}$$\Rightarrow$$\mathtt{respire,pond}$.\\
Ces règles font toutes partie du même R-ensemble noté
\textbf{R}($\mathtt{respire,pond,vole}$).

\subsection{Propriétés de la M-subsomption et du R-ensemble}

Soient $R_{1}$, $R_{2}$ et $R_{3}$ trois règles extraites
respectivement des concepts $C_{1}$, $C_{2}$ et $C_{3}$.
\begin{enumerate}
\item \textbf{transitivité} \label{trans}si $R_{1}$
$\sqsubseteq$$_{M}$ $R_{2}$ et $R_{2}$ $\sqsubseteq$$_{M}$ $R_{3}$
alors $R_{1}$ $\sqsubseteq$$_{M}$ $R_{3}$. En effet, puisque si
$C_{1}$ $\sqsubseteq$ $C_{2}$ et $C_{2}$ $\sqsubseteq$ $C_{3}$
alors $C_{1}$ $\sqsubseteq$ $C_{3}$ car la subsomption entre
concepts est transitive. \item \textbf{réflexivité} $R_{1}$
$\sqsubseteq$$_{M}$ $R_{1}$ car $C_{1}$ $\sqsubseteq$ $C_{1}$.
\label{refl} \item \textbf{anti-symétrie} si $R_{1}$
$\sqsubseteq$$_{M}$ $R_{2}$ et $R_{2}$ $\sqsubseteq$$_{M}$ $R_{1}$
alors $R_{1}$ et $R_{2}$ sont du même R-ensemble, car si $C_{1}$
$\sqsubseteq$ $C_{2}$ et $C_{2}$ $\sqsubseteq$ $C_{1}$ cela
implique que $C_{1}$ = $C_{2}$ et donc $R_{1}$ et $R_{2}$ sont
extraites du même concept dans le treillis. \label{anti}
\end{enumerate}

Les deux propriétés \ref{trans} et \ref{refl} définissent un
pré-ordre sur l'ensemble des règles et les trois propriétés
\ref{trans}, \ref{refl} et \ref{anti} définissent un ordre partiel
sur les R-ensembles.

\subsection{Expérimentation sur la base du "zoo"}

Dans cette section, nous présentons une expérimentation illustrant
la M-sub\-som\-ption sur une base de données réduite "Zoo"
\cite{Forsyth91} où les individus dénotent des animaux
($\mathtt{antilope, ours, sanglier,..}$) et les propriétés
($\mathtt{pond, respire, vole,...}$) sont non hiérarchisées. Cette
base de données compte 40 individus et 19 propriétés binaires.
Nous avons construit le treillis de Galois et extrait les règles
d'association à l'aide du logiciel Galicia \cite{Valtchev03}. Les
règles d'associations ont était extraites avec \emph{minsupp} =
0.3 et \emph{minconf} = 0.5. Nous avons obtenu 38 règles
partielles et 7 règles totales. Les règles extraites ont été
hiérarchisées selon la M-subsomption.

\begin {figure}[!ht]
\begin {center}
\scalebox{0.9}{\includegraphics[width=\textwidth] {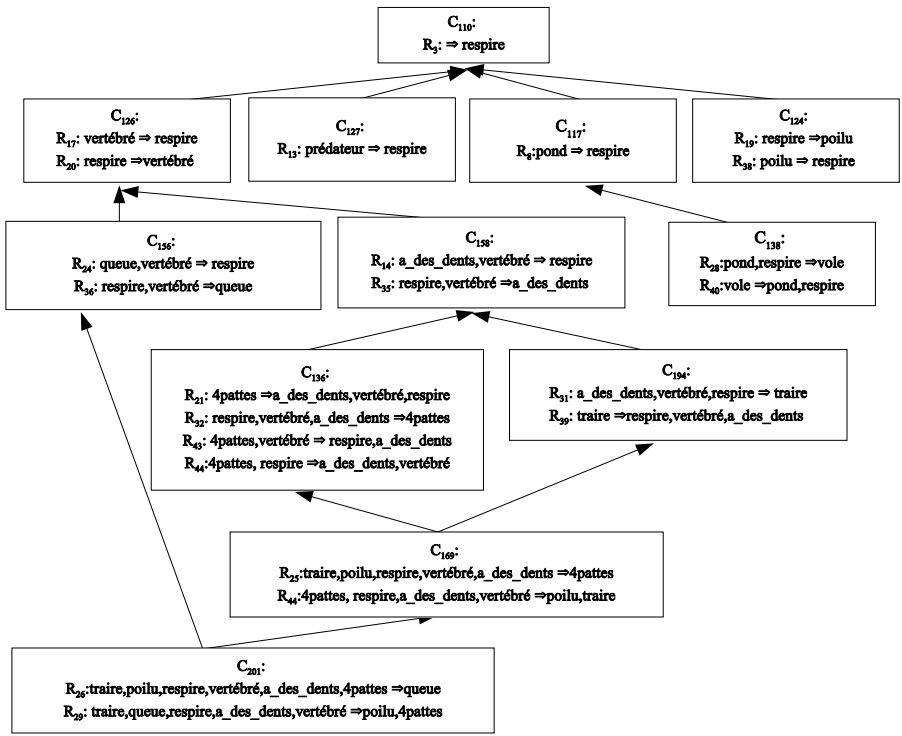}}
\end {center}
\caption{\textbf{Une partie de la hiérarchie des règles}}
\label{prop1}
\end{figure}

Une partie de la hiérarchie des règles est données par la figure
\ref{prop1}. Chaque concept du treillis $C_{i}$ (par exemple
$C_{117}$, $C_{127}$...) définit un R-ensemble et les liens entre
les R-ensembles sont les relations de subsomption existant entre
les règles des différents R-ensembles. Supposons que l'expert
prenne comme point de départ la règle $R_{3}$: $\Rightarrow$
$\mathtt{respire}$, qui veut dire l'ensemble des individus
possèdent la propriété $\mathtt{respire}$. Si cette règle lui
semble intéressante, mais trop générale, il peut rechercher des
règles plus spécifiques qui portent sur une population réduite.
Les concepts $C_{126}$, $C_{117}$, $C_{127}$ et $C_{124}$ sont
subsumés par le concept $C_{110}$, donc toutes les règles
extraites de ces concepts sont M-subsumées par la règle $R_{3}$.
Si l'expert veut prendre en compte la propriété
$\mathtt{vert\acute{e}br\acute{e}}$, il peut considèrer le concept
$C_{126}$, et toutes les règles valides issues de ce concept lui
sont proposées. Il peut réduire encore sa population ou augmenter
l'information contenue dans les règles en travaillant sur un motif
plus grand en ajoutant d'autres propriétés, comme :
$\texttt{a-des-dents}$ (concept $C_{158}$) ou $\texttt{queue}$
(concept $C_{156}$). En choisissant la propriété
$\texttt{a-des-dents}$, il accède à un R-ensemble de deux règles
$R_{14}$: $\texttt{a-des-dents}$,
$\mathtt{vert\acute{e}br\acute{e}}$ $\Rightarrow$
$\mathtt{respire}$ et $R_{35}$:
$\mathtt{respire}$,$\mathtt{vert\acute{e}br\acute{e}}$$\Rightarrow$
$\texttt{a-des-dents}$. S'il descend encore dans la hiérarchie
vers les concepts $C_{136}$ et $C_{194}$ et qu'il trouve que sa
population a été trop réduite (le support a trop diminué), il peut
s'arrêter à ce niveau de la hiérarchie et ne pas consulter les
règles plus spécifiques.

Maintenant, si l'analyste étudie la règle $R_{13}$:
$\mathtt{pr\acute{e}dateur}$ $\Rightarrow$ $\mathtt{respire}$, il
se rendra compte qu'il n'existe aucune propriété pouvant être
ajoutée au motif "$\mathtt{pr\acute{e}dateur}$,
$\mathtt{respire}$" pour avoir une règle valide, car il n'existe
aucun descendant du R-ensemble
\textbf{règles}($\mathtt{pr\acute{e}dateur}$, $\mathtt{respire}$)
dans la hiérarchie des règles.

Cette méthode de classification des règles est simple et ne
demande pas de calcul supplémentaire, la subsomption de règle
étant directement issue du treillis. Elle offre à l'analyste une
hiérarchie globale pour l'analyse des règles.

\section{Subsomption des règles avec un modèle terminologique}

La M-subsomption permie à l'expert d'avoir une vision globale et
structurée de l'ensemble des règles. Supposons à présent que
l'expert soit plus particulièrement intéressé par une règle et
qu'il dispose d'un modèle terminologique qui structure en une
hiérarchie l'ensemble des propriétés P. Nous définissons un second
type de subsomption qui permet de générer de nouvelles règles
généralisant la règle étudiée. Cette subsomption crée donc par
rapport au treillis global une structure hiérarchique locale dont
nous présentons les propriétés formelles.

En premier lieu, précisons ce que nous appelons modèle
terminologique. De façon analogue au modèle de connaissances
introduit dans la construction d'un treillis (Godin et al. 1995)
notre modèle terminologique est une hiérarchie de propriétés
$\mathcal{T}$ construite selon la relation \emph{Est-un}, définie
sur $\mathcal{T \times T}$. L'interprétation de A \emph{Est-un} B
signifie que si un individu possède A alors il possède B qu'on lui
rajoute car B n'est pas dans la base de données. La relation
\emph{Est-un} est réflexive, transitive et anti-symétrique, c'est
donc un ordre partiel. Les propriétés de l'ensemble P (du contexte
(I,P,$\mathcal{R}$) sont des feuilles pour la relation
\emph{Est-un}.

\subsection{La subsomption en programmation logique inductive}
\label{SPLI}

La programmation logique inductive \cite{Cornuejols01} réalise
l'apprentissage de formules de la logique des prédicats à partir
d'exemples et de contre-exemples. L'enjeu est de construire des
expressions logiques comportant des variables liées les unes aux
autres.

L'objectif de la PLI est la construction de formules logiques
incluant le plus d'exemp\-les, et le moins de contre-exemples
possibles. Notre objectif pour les règles est comparable. Nous
souhaitons engendrer une règle qui généralise une ou plusieurs
règles sans pour autant sur-généraliser et englober des
contre-exemples.

Il existe plusieurs types de formules en logique des prédicats et
celles qui nous intéressent sont les clauses, qui montrent une
certaine similitude avec règles d'association. Nous allons
rappeler la définition des clauses et nous inspirer de la
définition de la subsomption entre clauses pour calquer la
subsomption entre règles d'association.

\begin{definition}[Clause, théorie et subsomption relative à une théorie]\mbox{}

Une $\textbf{clause}$ est une formule de la logique des prédicats,
qui se compose d'une disjonction finie de littéraux dont toutes
les variables sont quantifiées universellement. Une clause s'écrit
:
$\neg$$B_{1}$$\vee$$\neg$$B_{2}$$\vee$...$\vee$$\neg$$B_{n}$$\vee$$A_{1}$$\vee$$A_{2}$....$\vee$$A_{m}$ ou encore en abrégé:\\
$B_{1}$,$B_{2}$,...,$B_{n}$$\rightarrow$$A_{1}$,$A_{2}$,...,$A_{m}$.

Une $\textbf{théorie}$ est un ensemble de clauses.

La clause $C_{1}$ \textbf{subsume} la clause $C_{2}$ relativement
à la théorie T si : de T $\wedge$ $C_{1}$ nous pouvons déduire
$C_{2}$, ce que nous notons : T $\wedge$ $C_{1}$ $\models$ $C_{2}$
ou $C_{1}$ $\models$$_{T}$ $C_{2}$.
\end{definition}

De ces définitions, nous dérivons la subsomption entre règles
d'association que nous nommons la H-subsomption, notée
$\sqsubseteq$$_{H}$. En premier lieu, nous introduisons la notion
d'ancêtre d'une propriété. Soient A, B deux propriétés du modèle
terminologique. Si A $(\emph{Est-un})^{*}$ B (la relation
\emph{Est-un} peut-etre appliquée plusieurs fois), alors il existe
un chemin dans la hiérarchie du modèle terminologique entre A et
B. Tout ancêtre de A est noté Â.

\begin {figure}[!h]
\begin {center}
\scalebox{0.22}{\includegraphics[width=\textwidth] {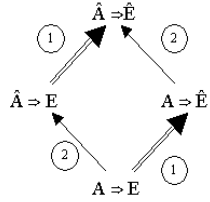}}
\end {center}
\caption{\textbf{Généralisation d'une règle}} \label{prophie}
\end{figure}

\begin{definition}[H-Subsomption des règles] \mbox{}
Soient deux règles \texttt{$R_{1}$: A $\Rightarrow$ B} et\\
\texttt{$R_{2}$: C $\Rightarrow$ D}, \texttt{$R_{1}$
$\sqsubseteq$$_{H}$ $R_{2}$} ssi C est un ancêtre de A et D est un
ancêtre de B.

\end{definition}

Nous nous refusons à garder dans la hiérarchie des règles, des
règles du type\\ \texttt{A $\Rightarrow$ Â} ou \texttt{Â
$\Rightarrow$ A}. En effet, ces règles ne sont pas informatives
(A$\cap$Â $\neq$ $\emptyset$)

L'idée de la H-subsomption a été partiellement reprise de
(\cite{Agrawal95}, \cite{Maedche00}), sur la généralisation de
règles d'association en s'appuyant sur une hiérarchie des
propriétés. Cependant au lieu de supprimer une règle du type A
$\Rightarrow$ B lorsqu'existe la règle Â $\Rightarrow$ B avec Â
l'ancêtre de A, nous avons défini la H-subsomption entre ces deux
règles et nous les avons gardées toutes les deux, car nous pensons
que la suppression de A $\Rightarrow$ B va entraîner une perte
d'information.

La figure \ref{prophie} montre la hiérarchie des règles qui
généralisent la règle \texttt{A $\Rightarrow$ E} telle que A et E
sont des ensembles de propriétés. Nous allons décrire comment
cette généralisation a été faite, sachant que pour un ensemble de
propriétés $P_{1}$ = $p_{1}$$p_{2}$...$p_{n}$ en remplaçant
$p_{i}$ tel que 1$\leq$i$\leq$n par $\mathtt{\widehat{p_{i}}}$
dans $P_{1}$ on obtient un ancêtre de $P_{1}$ noté
$\mathtt{\widehat{P_{1}}}$. Il y a deux types de généralisation
différentes notées dans la figure \ref{prophie} par \ref{1} et
\ref{2}:

\begin{dingautolist}{192}
        \item  Généralisation de la partie droite de la règle : Si  $\mathtt{A \Rightarrow E}$ alors  $\mathtt{A \Rightarrow \widehat{E}}$.

        La démonstration est immédiate en appliquant la propriété Prop2 de la section \ref{P2}. En effet Image(E)$\subseteq$
        Image(Ê),puisque Ê est plus général que E et donc\\ support($\widehat{E}$ $\sqcap$ A) $\geq$ support(E $\sqcap$
        A). La règle $\mathtt{A \Rightarrow \widehat{E}}$ est
valide car:
\[\vspace{-0.2cm}
 confiance(\mathtt{A \Rightarrow \widehat{E}}) =
\frac{support(\widehat{E} \sqcap A)}{support(A)} \geq
confiance(\mathtt{A \Rightarrow E}) = \frac{support(E \sqcap
A)}{support(A)}\vspace{-0.2cm}
\] \label{1}
        \item  Généralisation de la partie gauche de la règle : Si  $\mathtt{A \Rightarrow E}$ alors  $\mathtt{\widehat{A} \Rightarrow E}$ sous condition.

        Ce type de généralisation est de nature inductive. Nous considérons la règle  $\mathtt{A \Rightarrow \widehat{A}}$ comme étant
        une théorie. De $\mathtt{A \Rightarrow \widehat{A}}$ et de la règle  $\mathtt{\widehat{A} \Rightarrow E}$,
        nous pouvons déduire par transitivité des règles vu à la propriété Prop1 dans \ref{P1}, la règle  $\mathtt{A \Rightarrow E}$,
        et d'après la définition de la subsomption en PLI, nous
        pouvons déduire que la règle  $\mathtt{\widehat{A} \Rightarrow E}$ subsume la règle  $\mathtt{A \Rightarrow E}$,
        par rapport à la théorie  $\mathtt{A \Rightarrow \widehat{A}}$.
        Pour ce type de généralisation, le support( $\mathtt{\widehat{A} \Rightarrow E}$) $\geq$ support( $\mathtt{\widehat{A} \Rightarrow E}$) $\geq$ $\emph{minsupp}$,
        mais il faut vérifier la confiance( $\mathtt{\widehat{A} \Rightarrow E}$) pour que la nouvelle règle reste valide car le
        support de la partie gauche de la règle a augmenté ce qui
        peut entraîner une sur-généralisation et donc le fait
        d'englober trop de contre-exemples. \label{2}

\end{dingautolist}

\subsection{Propriétés de la H-subsomption} Soient $\mathtt{R_{1}}$ : $\mathtt{A \Rightarrow B}$,
$\mathtt{R_{2}}$ : $\mathtt{C \Rightarrow D}$ et $\mathtt{R_{3}}$
: $\mathtt{E \Rightarrow F}$
\begin{enumerate}
    \item transitivité : si $\mathtt{R_{1}}$ $\sqsubseteq$$_{H}$ $\mathtt{R_{2}}$ et $\mathtt{R_{2}}$
    $\sqsubseteq$$_{H}$$\mathtt{R_{3}}$ alors : $\mathtt{R_{1}}$ $\sqsubseteq$$_{H}$
    $\mathtt{R_{3}}$, en effet car si C = Â\\ et E = $\widehat{C}$ alors E = Â et D =
    $\widehat{B}$ et F = Ê alors F = $\widehat{B}$. \label{a}
    \item réflexivité : comme nous considérons que chaque propriété est son propres ancêtre alors : $\mathtt{R_{1}}$ $\sqsubseteq$$_{H}$
    $\mathtt{R_{1}}$. \label{b}
    \item anti-symétrie : si $\mathtt{R_{1}}$ $\sqsubseteq$$_{H}$ $\mathtt{R_{2}}$ et $\mathtt{R_{2}}$ $\sqsubseteq$$_{H}$ $\mathtt{R_{1}}$ alors $\mathtt{R_{1}}$ =
    $\mathtt{R_{2}}$, car si A = $\widehat{C}$ et C = Â alors A = C et B =
    $\widehat{D}$
    et D = $\widehat{B}$ alors D = B. \label{c}
\end{enumerate}
    Ces trois propriétés définissent un
ordre partiel.

La hiérarchie des règles peut ne pas être un treillis complet car
la borne supérieure peut ne pas exister. Ceci est du à l'exclusion
des règles du type $\mathtt{A \Rightarrow \widehat{A}}$ et
$\mathtt{\widehat{A} \Rightarrow A}$ et par le fait que pour
certaines généralisations nous devons contrôler la confiance.

\subsection{Expérimentation sur des règles avec modèle terminologique}

Nous avons expérimenté la H-subsomption sur une base de données de
6 individus et de 6 propriétés. On suppose que cette base a été
créée par un professeur qui voudrait savoir quelles sont les
grandes tendances dans le choix des modules. Le tableau de cette
base est présenté dans la table \ref{E2T1}.

\begin {figure}[htbp]
\begin{center}
\scalebox{0.6}{\includegraphics[width=\textwidth] {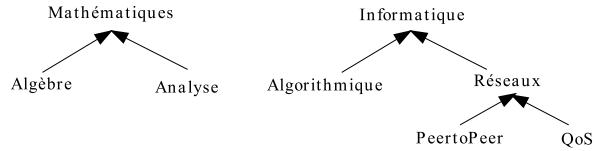}}
\caption{\textbf{Hiérarchie des propriétés}} \label{E2F1}
\end{center}
\end{figure}

\begin{table}[!ht]
\begin{center}
\scriptsize
\begin{tabular}{|c|c|c|c|c|c|c|}
  \hline
  \scriptsize{R} & \scriptsize{Algèbre} & \scriptsize{Algorithmique} & \scriptsize{Probabilité} & \scriptsize{QoS} & \scriptsize{PeertoPeer} & \scriptsize{Biologie} \\
  \hline
  $I_{1}$ & 1 & 1 & 1 & 1 & 1 & 0 \\
  \hline
  $I_{2}$ & 1 & 1 & 0 & 0 & 0 & 1 \\
  \hline
  $I_{3}$ & 0 & 0 & 1 & 0 & 0 & 0 \\
  \hline
  $I_{4}$ & 0 & 1 & 1 & 0 & 1 & 0 \\
  \hline
  $I_{5}$ & 1 & 1 & 1 & 1 & 0 & 0 \\
  \hline
  $I_{6}$ & 0 & 0 & 0 & 0 & 1 & 1 \\
  \hline
\end{tabular}
\normalsize
\end{center}
\caption{\textbf{Représentation en tableau de la relation R}}
\label{E2T1}
\end{table}

Nous avons fixé $\emph{minsupp}$ = 0.5 et $\emph{minconf}$ = 0.5.
Nous avons obtenu en appliquant un algorithme d'extraction de
règles à partir de motifs fréquents 9 règles partielles $P_{i}$ et
1 règle totale $T_{0}$, qui sont présentées dans le tableau
\ref{E2T2}.

\begin{table}[!ht]

\begin{center}
\scriptsize
\begin{tabular}{|c|c|c|c||c|c|c|c|}
  \hline
  &  Règles & Sup. & Conf. & & Règles & Sup. & Conf.\\
  \hline
  $P_{0}$ & $\Rightarrow$ PeertoPeer & 0.5 & 0.5 & $P_{5}$ & Algorithmique $\Rightarrow$ Probabilité & 0.5 & 0.75 \\
  \hline
  $P_{1}$ & $\Rightarrow$ Probabilité & 0.66 & 0.66 & $P_{6}$ & $\Rightarrow$ Probabilité,Algorithmique  & 0.5 & 0.5 \\
  \hline
  $P_{2}$ & $\Rightarrow$ Algorithmique & 0.66 & 0.66 & $P_{7}$ & Algorithmique $\Rightarrow$ Algèbre & 0.5 & 0.75 \\
  \hline
  $P_{3}$ & $\Rightarrow$ Algèbre & 0.5 & 0.5 & $P_{8}$ & $\Rightarrow$ Algorithmique,Algèbre & 0.5 & 0.5 \\
  \hline
  $P_{4}$ & Proba $\Rightarrow$ Algorithmique & 0.5 & 0.75 & $T_{0}$ & Algèbre $\Rightarrow$ Algorithmique & 0.5 & 1 \\
  \hline
\end{tabular}
\normalsize
\end{center}

\caption{\textbf{Les règles extraites du tableau de \ref{E2T1}}}
\label{E2T2}
\end{table}

\begin {figure}[!h]
\begin {center}
\scalebox{0.74}{\includegraphics[width=\textwidth] {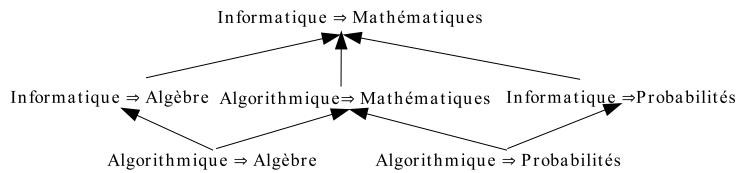}}
\end {center}
\caption{\textbf{Hiérarchisation des règles P5 et P7}}
\label{E2F4}
\end{figure}

Puis nous avons généralisé les règles pour lesquelles la partie
droite ou gauche est composée d'une propriété ayant un ancêtre
dans le modèle terminologique \ref{E2F1}. Pour les généralisations
de type \ref{2}, nous avons contrôlé la confiance. Nous illustrons
par la figure \ref{E2F4} la hiérarchie de règles construite à
partir des deux règles : P7 : $\mathtt{Algorithmique}$
$\Rightarrow$ $\mathtt{Alg\grave{e}bre}$ et P4 :
$\mathtt{Algorithmique}$
$\Rightarrow$ $\mathtt{Probabilit\acute{e}}$, la hiérarchie résultante est présentée dans la figure \ref{E2F4}\\

Supposons que le professeur cherche à caractériser le parcours des
étudiants et qu'il considère la règle $\mathtt{informatique}$
$\Rightarrow$ $\mathtt{Alg\grave{e}bre}$. Il peut chercher des
règles plus précises pour savoir s'il y a des sous-domaines de
l'information qui sont plus particulièrement concernés. Ainsi, il
accède à la règle $\mathtt{Algorithme}$ $\Rightarrow$
$\mathtt{Alg\grave{e}bre}$. \`A l'inverse, s'il veut une vision
plus générale de la règle $\mathtt{informatique}$ $\Rightarrow$
$\mathtt{Alg\grave{e}bre}$, alors il accède à la règle
$\mathtt{informatique}$ $\Rightarrow$
$\mathtt{Math\acute{e}matique}$ à travers un processus d'induction
validé par le contrôle de confiance. De cet exemple nous pouvons
déduire que la hiérarchie de généralisation peut être partagée par
plusieurs règles.

Ainsi, la subsomption entre règles avec des propriétés
hiérarchisées permet d'exploiter un modèle terminologique et
d'utiliser cette hiérarchie des propriétés pour généraliser les
règles d'association.

\section{Conclusion}

Lorsque le nombre de règles d'association est important,
l'analyste cherche à trouver des liens entre ces règles pour
pouvoir en déduire les connaissances connues dans son domaine, et
éventuellement de nouvelles connaissances. Nous avons ajouté la
classification des règles d'association dans l'étape de fouille de
données pour faciliter le travail de l'analyste lors de
l'évaluation et de l'interprétation des règles extraites. Nous lui
fournissons une hiérarchie de règles d'après les propriétés qui
les composent, qui lui permet de faire ressortir les liens dont il
a besoin.

Les règles d'association ayant des propriétés non hiérarchisées
sont classifiées dès qu'elles sont extraites du treillis de
Galois. De ce fait cette classification ne demande pas une étape
supplémentaire et offre plusieurs avantages à l'analyste tels que
le fait de redéfinir un support minimal s'il trouve que sa
population a trop été réduite dans le bas de la hiérarchie, et de
voir toutes les règles valides qui sont extraites du même concept
(les règles du même R-ensemble).

La structuration des règles d'association avec des propriétés
hiérarchisées permet de tenir compte des liens entre les
différentes propriétés et de pouvoir généraliser l'une des deux
parties de la règle.

La classification des règles d'association dans le cas d'une base
de textes sert à relier les textes entres eux. Dans le cas des
règles avec des propriétés hiérarchisées, cette relation entre les
textes peut être interprétée comme le fait qu'un texte mentionne
des termes plus spécifiques qu'un autre, ce qui peut aider
l'expert à classifier ces textes.
\section*{Références}

\nocite{*}

\bibliographystyle{apalike}
\bibliography{biblio}

\section*{Summary}

Extraction of association rules is widely used as a data mining
method. However, one of the limit of this approach comes from the
large number of extracted rules and the difficulty for a human
expert to deal with the totality of these rules. We propose to
solve this problem by structuring the set of rules into hierarchy.
The expert can then therefore explore the rules, access from one
rule to another one more general when we raise up in the
hierarchy, and in other hand, or a more specific rules.

Rules are structured at two levels. The global level aims at
building a hierarchy from the set of rules extracted. Thus we
define a first type of rule-subsomption relying on Galois
lattices. The second level consists in a local and more detailed
analysis of each rule. It generate for a given rule a set of
generalization rules structured into a local hierarchy. This leads
to the definition of a second type of subsomption. This
subsomption comes from inductive logic programming and integrates
a terminological model.

\end{document}